# Capping of Nanoparticles: An Alternative Approach for Reducing Nanoparticle Toxicity in Plants

Ritika Bhatt[1,2], Ankit Goyal[3]*, Sumita Kachhwaha[2], S.L. Kothari[4]

[1] Kanoria PG Mahila Mahavidhyalaya, Jaipur, India-302018
[2] Department of Botany, University of Rajasthan, Jaipur, India-302018
[3] Institute of Physics, University of Amsterdam, Science Park 904, Amsterdam, The Netherlands
[4] Institute of Biotechnology, Amity University Rajasthan, Jaipur, India

*Corresponding author- a.goyal@uva.nl

## Abstract

The present era has witnessed a new dawn in technology innovations with the entry and use of nanomaterials in the industries and in the products used and to be used in day-to-day life creating a huge possibility of ending up in the food chain. Several studies in the past has highlighted toxicity of nanomaterials due to their size. However, we cannot stop technological advancements provided by the nanomaterials fulfilling human needs but can find a solution to the toxicity of the nanomaterials for a better future and safe environment. In this study, we propose capping of nanomaterials to reduce the toxicity without compromising their functionality. Capping of the nanomaterials is used to passivate nanomaterials but the same capping also helps in the reduction of surface reactivity leading to low toxicity. We studied phytotoxicity in the presence of one of the most extensively used metal nanoparticles (copper nanoparticles) on *Eleusine corcana* G. (finger millet) and *Paspalum scrobiculatum* L. (Kodo millet). Copper nanoparticles were synthesized by the hydrometallurgical methods. Ethylenediaminetetraacetic acid (EDTA) was used to cap the nanoparticles during the synthesis. *In vitro* studies results showed that the toxicity of copper nanoparticles is significantly reduced after capping. Anti-bacterial activity studies showed no change in efficacy of copper nanoparticles after capping. This study highlights the use of capping to reduce the toxicity of nanomaterials without sacrificing their required applicability.

## Introduction

We are living in the era of nanotechnology, and with the development of new characterization and fabrication technologies, new nanotechnology-based products are constantly developing. Recent advances in semiconductors, polymers, metallurgy etc[1-2] has enabled the technology to reach to the masses which increases the interaction of nanomaterials with the living organisms and our ecosystem in applications like water purification and medicine etc[3-4]. Due to immense use of nanomaterials in consumer products and poor disposal of such materials, the engineered nanoparticles (ENP's) will eventually exposed into terrestrial and aquatic ecosystems, with their fate being still unknown to us[5]. Plants are key components of all ecosystems, and ENP intrusion in their tissues is inevitable due to efficient uptake and accumulation by roots[6].

Metal nanoparticles are under intense research, and their potential is exploited for development of many products. Many scientists have studied the toxicological effect of several nanomaterials on plants and animals[6-9]. However, only few efforts are made on the study of the effect of metallic (and their oxides) nanoparticles, majorly including silver and zinc ENPs only on plant germination and growth[10-11]. The effect of rare earth oxide nanoparticles on root elongation of plants was studied[12]. The effects of five types of nanoparticles (multi-walled carbon nanotube, aluminum, alumina, zinc, and zinc oxide) on seed germination and root growth of six higher plant species (radish, rape, ryegrass, lettuce, corn, and cucumber) was also reported[13]. Another group studied the effect of biologically synthesized silver nanoparticles on *Bacopa monnieri* (Linn.) Wettst plant's growth metabolism[14]. Accumulation of zinc, copper, and cerium in carrot (*Daucus carota*) exposed to metal oxide nanoparticles, and metal ions were observed[15]. Toxic effects induced by the NPs are studied by several scientists with almost similar trends but no efforts have been made for reducing this toxicity caused by the size of NPs in the plants and ultimately to the environment.

Millets are small-seeded cereals which serve as excellent alternates as staple food in having equivalent nutritional competence with advantage of growing with minimum inputs[16]. They are major sources of food, feed, and fodder for millions in the third world countries of Asia and Africa and thereby demand an important place on the world food map. Millets are chosen as model plants in our study because these are underutilized grains which play a major role in the food security of millions of people in South Asia and Africa and this study was done in the state of Rajasthan in India which has huge copper mines and millet farming. Millets would play an important role in coming years of drastic climate change as they have wide genetic adaptation to diverse climates and have potential to thrive with low inputs.

Copper nanoparticles were chosen as model NPs, due to 1) Copper is considered a less-toxic element in general 2) copper nanomaterials have immense potential in several applications like anti-bacterial activity, thermal conductivity, high strength metals and alloys and EMI shielding. It is envisaged that a Cu-based pesticide era has emerged (Rai et al 2018) therefore there is a large possibility of its higher interaction with the ecosystem. Copper is considered as third most necessary metal in terms of service and widespread applications to human race (Liu et al 2018,). Lots of different fabrication methods are developed to synthesize copper nanoparticles[18-20]. However, we choose an industrial scale synthesis method generally used to extract copper from copper sulphate waste, making the study a greener effort for sustainable environment.

In this report, we propose a widely used capping of nanomaterials method (used to enhance stability of nanomaterials) to reduce environmental toxicity. Copper nanoparticles with and without capping are synthesized by hydrometallurgical methods. The effect of nanoparticles of the growth of millets are studied *in-vitro*. Anti-bacterial activity of nanoparticles were studied to see effect of capping on the applicability of nanomaterials.

**Results & Discussion**

**Particle size measurement**

Transmission electron microscopy (TEM) was used to determine average particle size of synthesized copper nanoparticles (Cu NPs). Fig. 1 (a-b) is showing TEM images of un-coated and coated CuNPs.

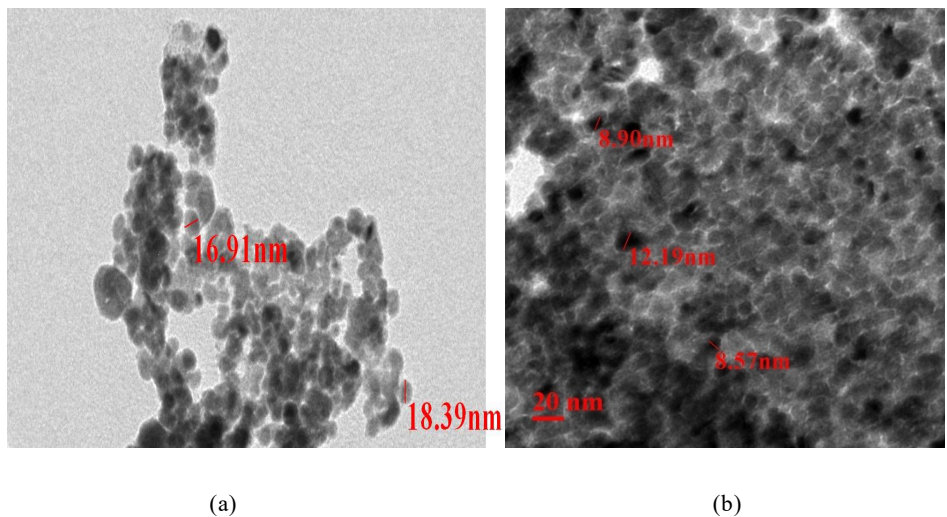

(a)　　　　　　　　　　　　　　(b)

Fig 1: TEM micrographs of Cu NPs a) Un-coated b) Coated

ImageJ software tool was used to measure particle size. It is clear from the micrographs that the particle size of both coated and uncoated copper particles is in the nanometer range. The average particle size of uncoated and coated CuNPs is ~ 8 nm and ~11 nm, respectively. A little higher particle size could be due to capping layers on the surface of the CuNPs. Agglomerated particles are visible in the micrograph of coated CuNPs because of the presence of EDTA capping layers.

**XRD Studies**

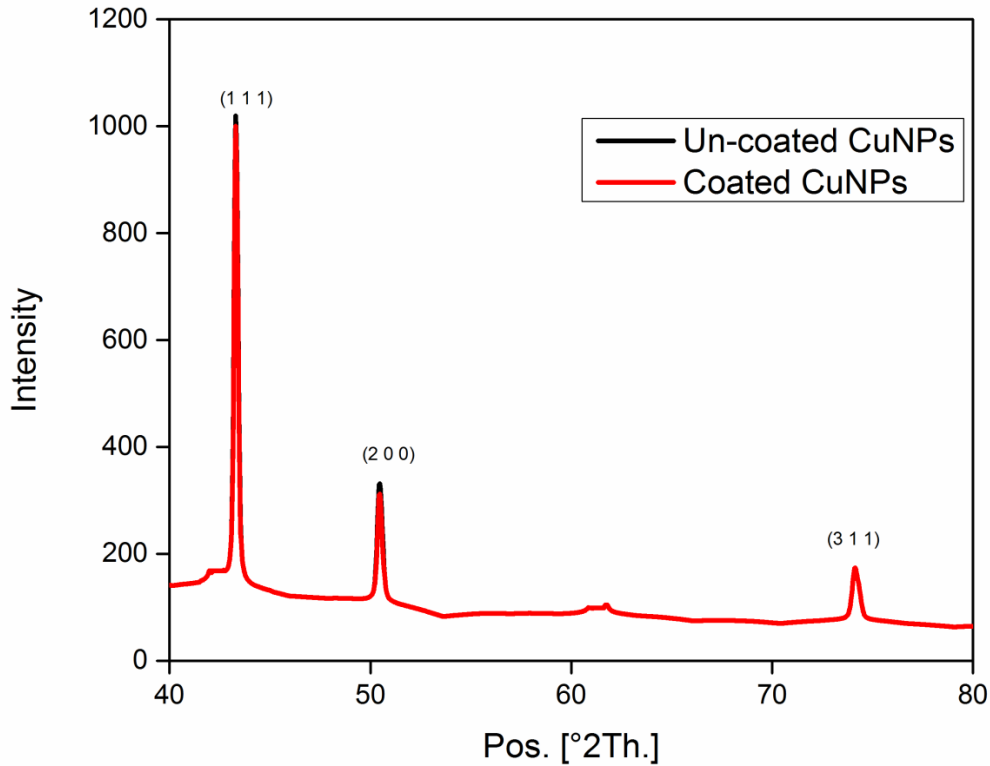

Fig. 2: XRD patterns of Copper nanoparticles. Red curve is for coated CuNPs and the Black is for Un-coated CuNPs. Due to overlapping of the patterns, the un-coated curve is a little bit visible at <111> and <200> planes.

P'analytical system diffractometer is used for XRD analysis of the samples. Fig. 2(a-b) shows XRD patterns of copper and uncoated CuNPs. XRD patterns have been obtained in 2θ range of 40º–80° with Cu Kα radiation source (λ=1.5406 Å). The X-ray tube was operated at 45 kV and 40 mA. XRD pattern is compared with JCPDE standard database [ref], and for both coated and un-coated nanoparticles, it shows polycrystalline phase with calculated h, k, l indices (111), (200), and (311) corresponding to peak positions 43.3273, 50.3961 and 74.1423. There was no change in the phase and crystallite size of the coated NPs even after capping.

The crystallite size of coated and uncoated CuNPs is approximated using Williamson-Hall relationship as in Eq. (1)

$$\beta_{tot} \cos\theta = C_\varepsilon \sin\theta + K\lambda/L \ldots\ldots\ldots\ldots\ldots(1)$$

Where, λ is wavelength of the X-rays used, $\beta_{tot}$ is full width at half maximum of the peak, K is a constant (K = 0.95 used for near circular samples), and θ is corresponding Bragg's angle. The crystallite size is calculated by size component (Kλ/L) which is equal to the intercept on the y-axis ($\beta_{tot} \cos\theta$). The approximated crystallite size is ~10 nm. The results of XRD also suggest nanocrystalline CuNPs as observed in TEM.

**Influence of nanoparticles on plant morphology**

Germination percentage of PR-202 was almost equivalent to that of control for all treatments while it was reduced for GPUK-3 (Table 2). The findings are like the earlier studies where different metal nanoparticles drastically reduced the germination percentage in wide variety of plants as of iron on rye grass, barley and flax[26], cerium oxide on maize, tomato, cucumber and soybean[27] and zinc oxide on corn[28]. Germination index was measured by calculating both the total and mean germination rate of seeds after specific time duration. Germination index was decreased in Kodo millet and more significantly upon uncoated nanoparticles exposure. The plausible reason could be that nanoparticles may be accumulated in the actively growing tissues especially in meristem, disrupting the regular growth mechanism and allowing the cells to choose alternate growth pathways to mitigate the toxicity effects causing decrease in germination index [ref]. We did not observed this difference in Finger millet and the reason is not known to us at the moment.

Table 2: Germination parameters for two millet species under different NP exposure

|  | Germination % | | Relative germination rate (d/e) | | Relative salt injury (e-d/e) | | Germination index | |
|---|---|---|---|---|---|---|---|---|
|  | Finger | Kodo | Finger | Kodo | Finger | Kodo | Finger | Kodo |
| Control | 100 | 87.5 |  |  |  |  | 4 | 3.5 |
| Coated NPs | 95 | 80 | 0.95 | 0.91 | 99.05 | 86.58 | 3.8 | 3.2 |
| Uncoated NPs | 95 | 52.5 | 0.95 | 0.6 | 99.05 | 86.9 | 3.8 | 2.1 |

The Fig. 3 (a-b) shows length measurements of *in vitro* grown finger millet and Kodo millet, respectively. After 15 days of inoculation, both the plantlets showed profuse growth in coated NP medium with no signs of toxicity or stress, as compared to the control plantlets. Reduced growth was observed in un-coated NPs treated plantlets suggest CuNPs increased the inhibition of uptake of other nutrients, may be possible replacement of $Fe^{+2}$ ions with access $Cu^{+2}$ ions thus reducing overall plant growth to critically minimum. The coated CuNPs might have affected the ion channels in the plants which are responsible for nutrient uptake and transportation, leading to marginal growth in uncoated ones.( any reference?)

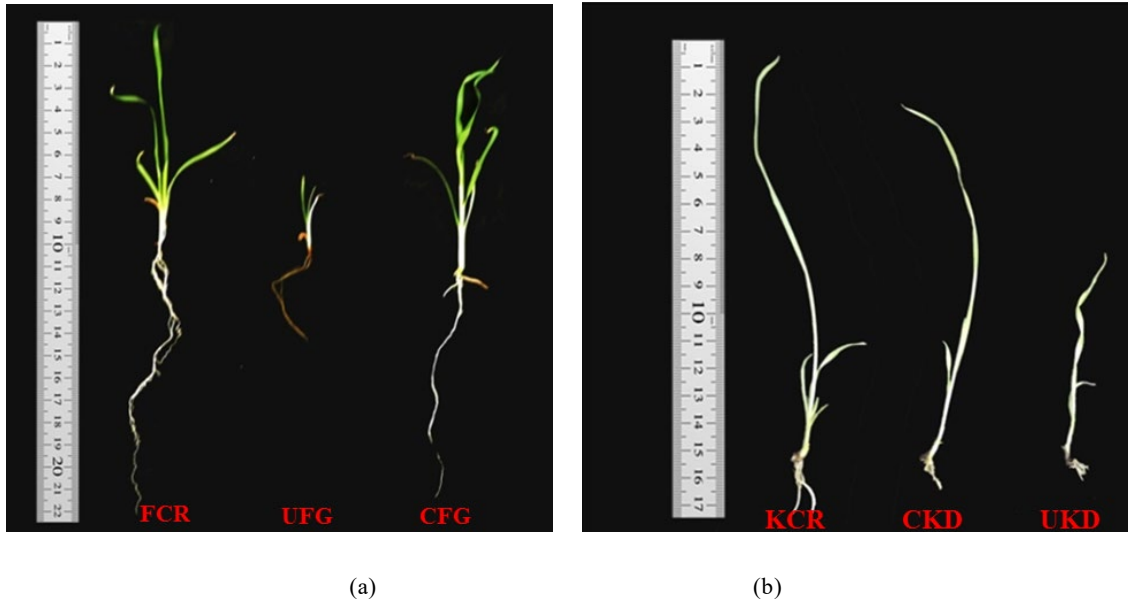

(a) (b)

Fig 3: Length measurements of a) finger millet; b) Kodo millet. Scale in is cm.

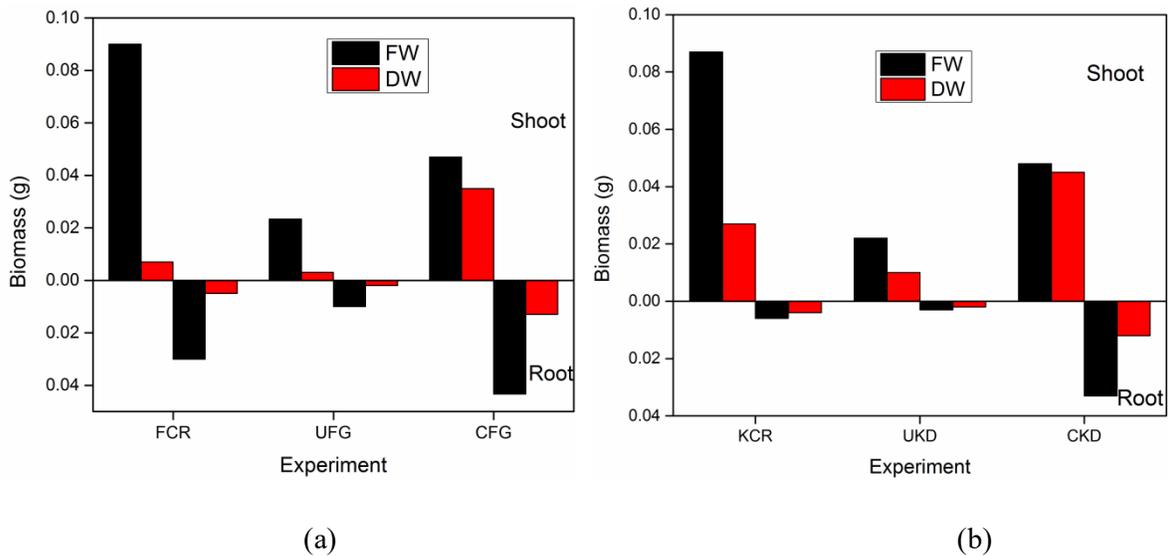

(a) (b)

Fig. 4: Biomass graphs of a) Finger millet b) Kodo millet. The black bars show fresh weight and red bars is for dry weight of plants.

The fig. 4 (a-b) shows the biomass charts for both the millet species. The treatment by coated and uncoated CuNPs reduced the fresh weight (FW) in comparison to control due to lesser growth which has shown similar trend of results. However, dry weight of CFG and CKD plantlets was higher than controls. Stampoulis et al 2009 have reported 90 % reduction in biomass in *Curcurbita pepo* in nanoparticles treated plants which is in line with our observations as we see ~70 % reduction in biomass in uncoated nanoparticles treated samples. Coating of nanoparticles helped to reduce this reduction in biomass to ~40 %.

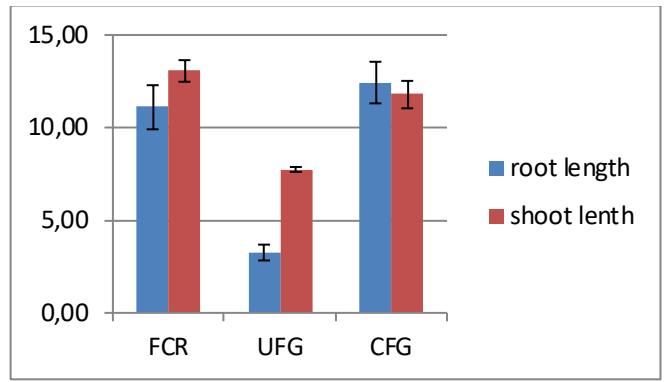

Fig 5: Average of length of shoots and roots in Finger millet

The results of average of individual root and shoot length can be seen in the Fig. 5 for Finger millet. There was no detrimental effect of coated CuNPs treatment on shoot and root length rather a little increase in the root length was observed on an average in comparison to control that is in error limits. Uncoated CuNPs reduced the growth of roots and shoots in the plants that could be due to presence of access $Cu^{+2}$ ions, affecting the functioning of the growth mechanism in root and shoots of plants.( any reference?)

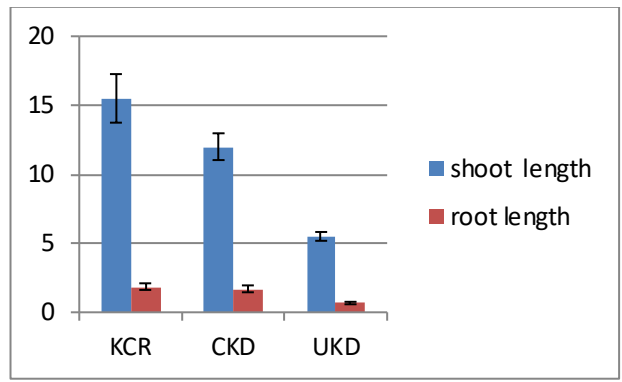

Fig 6: Average of length of shoots and roots in Kodo millet

In the case of Kodo millet, significant reduction in shoot and root length after uncoated CuNPs treatment as shown in Fig. 6 like what we observed in Finger millet. Here, the growth was reduced in the plantlets after coated CuNPs treatment too. The relative growth was higher in the plantlets treated by coated CuNPs. The results are similar to the results obtained by Atha et al 2012 where they have found naked CuNPs significantly affect the root development in *Raphanussativus* and *Lolium*. It also suggests the concentration of metal ions required for optimal growth is different for every species and thus need for more deep studies in various plant species is required.

The effects on root and shoot lengths were found to differ widely in several plant species like decrease in the ratio was seen for zucchini plants[30] but an increase was observed in lettuce plants[31] upon treatment with CuNP as compared to control. Cu NPs had no effect on shoot length but it decreased shoot biomass in comparison to control in plants of Oregano grown in soil augumented with Cu-NPs( Du et al 2018)

## Influence of Cu nanoparticles on plant's biochemical parameters

### Enzyme assay

The immediate molecular response by any plant species under stress is the production of reactive oxygen species which in turn execute the signal transduction pathways by acting as secondary messengers to relay the information further to trigger the defense mechanism of the plant[32]. Different cell organelles like mitochondria, peroxisome, etc. detoxify these ROS by the outburst of antioxidant enzymes like superoxide dismutase, catalase, peroxidase[33].

Normally it has been reported several times that introduction of heavy metal ions or particles like silver and gold (above certain concentration) increase the enzymatic activity like POX and CAT activity in the test plants due to increment in the ROC in the tissues [references]. But here, reduction in POX activity was seen in both the cases and more pronounced in the case of coated CuNPs (Fig 7). This is due to combine effect of polymer capping of nanoparticles which reduced the speed of oxidation of metal particles into ions causing less ROC in the tissues and lesser concentration of metal nanoparticles in the medium.

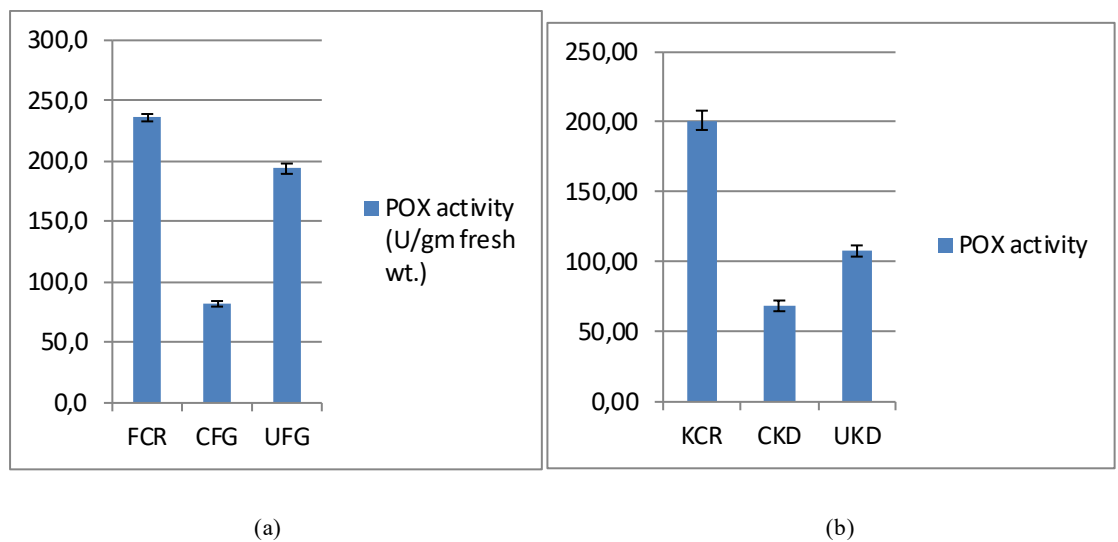

Fig 7: POX activity shown by: a) Finger millet ; b) Kodo millet

Oxidation of nanoparticles results in increase in the concentration of the metal ions and reduction of ions, increases the concentration of metal nanoparticles in the cytoplasm[34]. Both oxidation and reduction agents are present in abundance in the plants which regulate the concentration of ions and particles in the plant[35]. Copper nanoparticles coated with polymers are stabilized and released slowly thereby reducing their participation in redox reactions. The uptake and incorporation of nanoparticles into the plant cells have already been discussed by many researchers[36-39]. CuNPs is reported to elevate the levels of catalase in roots of lettuce exposed to NPs hydroponically ( Trujiilo-Reyes et al 2014) and it was found that in roots of lettuce exposed to Cu/CuO NPs, the activity of catalase (CAT) increased, while that of ascorbate peroxidase (APX) decreased.

## Chlorophyll content

Copper element plays important role in metabolism of macromolecules as well is directly involved in ETS of photosynthesis (Jain et al 2014). Fig. 8 is showing effect of coated and uncoated CuNPs on the chlorophyll content of the plant.

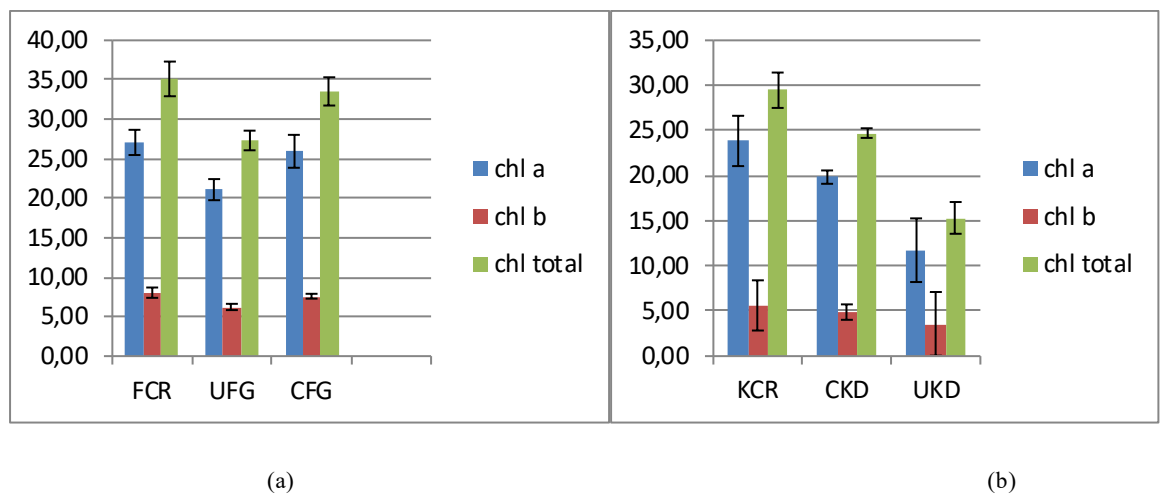

Fig 8: a) Chlorophyll content in finger millet, b) Chlorophyll content inKodo millet

The decrease in chlorophyll content concord with the earlier study done on maize plant, where chl a level was found to be decreased[41], also in Cucumis plants where CuNP treated plants exhibited a decrease in Chlorophyll a and b contents (Mosa et al 2018). Reduced chlorophyll content could also be due to dampened protochlorophyllidereductase activity, which contributes to chlorophyll synthesis, and enhanced activity of chlorophyllase, a chlorophyll degrading enzyme under stress conditions, caused by Cu toxicity ( Eser and Aydemir, 2016). The uncoated CuNPs have released higher concentration of copper ions, causing increase in activity of chlorophyllase and also hindered the electron mobility thus affected chlorophyll activities (Apodaca et al 2017). The significant difference in the content of both types of chlorophyll can be seen in coated and uncoated CuNPs suggesting reduction in the toxicity caused by these nanoparticles in presence of capping agents.

Mechanism of toxicity of NPs in the plant is not fully understood and nor properly explored till yet, but few attempts by studying release of metal ions, generation of reactive oxygen intermediates (ROIs), and oxidative stress are helpful for explaining the toxicity caused.[42-44]. The size of nanoparticles and their possible interaction with different chemical and physical reagents in the plants can be the possible reason. Researchers also compared the toxicity of ZnO NPs with that of $Zn^{2+}$ and found that toxicity of ZnO NPs might be partly related to the solubility of the NPs[44]. This can be true in this case also as solubility of metal NPs in water is zero which causes their direct interference in any interaction of plants to the nutrients and water. The agglomeration of particles is also there which causes blockade in the cell pores and hinder nutrient absorption. The release of ions also changes the concentration of ions in the interacting media which lead to polarity change. This affects flow of catalysts, formation of proteins and enzymes even messengers. Root tips and hairs can secrete large amounts of mucilage, coating the root surface. This mucilage is a highly hydrated polysaccharide, probably a pectin substance, which might contribute to the adsorption of NPs on the root surface.

Presence of gelatin during chemical synthesis increased stability of synthesized CuNPs (Satyvaldiev et al 2018). The results of present study clearly indicate the better performance of plants exposed to coated CuNPs than uncoated CuNPs which gives a boost to the theory of control release of metal ions due to presence of the capping agent in copper nanoparticles which make them less toxic than uncoated counterparts. This also suggests possible control of size of nanoparticles and use of nanoparticles in lesser toxic way. The release of metal nanoparticles coated with biocompatible polymers can reduce their toxicity vastly. It was reported that sledges contain elevated levels of metal containing nanoparticles (Tou et al 2017).

## 5. Conclusions

Polymer capping has minimized the toxic behavior of copper nanomaterials to an extent in food plants. The results show a new way of reducing the toxicity caused by applied nanomaterials in the environment.

## Acknowledgements

The authors gratefully acknowledge DRS Phase II, Department of Botany, DBT- BIF and DST PURSE, University of Rajasthan for providing necessary facilities.